\documentclass[
notoc,nohyper]{JHEP} 

\usepackage[psamsfonts]{amsfonts}

\newcommand{\Gammait}{{\mit\Gamma}}
\newcommand{\Lambdait}{{\mit\Lambda}}
\newcommand{\tr}{\mathop{\rm tr}\nolimits}

\newcommand{\SU}{\mathop{\rm SU}}
\newcommand{\U}{\mathop{\rm {}U}}
\newcommand{\rmd}{{\rm d}}

\newcommand\fverb{\setbox\pippobox=\hbox\bgroup\verb}
\newcommand\fverbdo{\egroup\medskip\noindent%
                        \fbox{\unhbox\pippobox}\ }
\newcommand\fverbit{\egroup\item[\fbox{\unhbox\pippobox}]}
\newbox\pippobox

\title{
Axial anomaly in the reduced model:\\ Higher representations}

\author{Teruaki Inagaki\\
Graduate School of Science and Engineering, Ibaraki University, Mito 310-8512,
Japan\\
E-mail: \email{teruaki@serra.sci.ibaraki.ac.jp}}
\author{Yoshio Kikukawa\\
Department of Physics, Nagoya University, Nagoya 464-8602, Japan\\
E-mail: \email{kikukawa@eken.phys.nagoya-u.ac.jp}}
\author{Hiroshi Suzuki\\
Department of Mathematical Sciences, Ibaraki University, Mito 310-8512, Japan\\
E-mail: \email{hsuzuki@mx.ibaraki.ac.jp}}
\received{\today}               
\accepted{\today}               

\preprint{IU-MSTP/56\\DPNU-03-12\\\heplat{0305011}}     

\abstract{
The axial anomaly arising from the fermion sector of $\U(N)$ or $\SU(N)$
reduced model is studied under a certain restriction of gauge field
configurations (the ``$\U(1)$ embedding'' with $N=L^d$). We use the
overlap-Dirac operator and consider how the anomaly changes as a function of
a gauge-group representation of the fermion. A simple argument shows that the
anomaly vanishes for an irreducible representation expressed by a Young tableau
whose number of boxes is a multiple of~$L^2$ (such as the adjoint
representation) and for a tensor-product of them. We also evaluate the anomaly
for general gauge-group representations in the large~$N$ limit. The large
$N$ limit exhibits expected algebraic properties as the axial anomaly.
Nevertheless, when the gauge group is $\SU(N)$, it does not have a structure
such as the trace of a product of traceless gauge-group generators which is
expected from the corresponding gauge field theory.}

\begin{document} 

\maketitle 

\section{Introduction}
It has been unclear for long time how to define the topological charge in the
reduced model for large~$N$ QCD~\cite{Eguchi:1982nm}--\cite{Das:1984nb}. One
may try to define the topological charge in a ``fermionic way'' through the
index theorem. However, since the reduced model is given by a zero-volume limit
of a field theory, it is a system of finite degrees of freedom as long as $N$
is finite even very large. It is then obvious that one cannot have the axial
anomaly, unless a certain source of an explicit breaking of axial symmetry is
introduced from an onset. Then one may ask: What is a good way to simulate the
(quantum) axial symmetry breaking in a system of finite degrees of freedom?

Recently, motivated by a success of the overlap-Dirac
operator~\cite{Neuberger:1998fp} in lattice gauge theory (which is also a
finite system when the lattice size is finite), authors of a
paper~\cite{Kiskis:2002gr} proposed a use of the overlap-Dirac operator in the
quenched reduced model~\cite{Bhanot:1982sh}. The overlap-Dirac operator
satisfies the Ginsparg-Wilson relation~\cite{Ginsparg:1982bj} and this relation
ensures remarkable properties concerning the chiral symmetry. For example, an
index relation which is analogous to that in the continuum holds even with
finite degrees of freedom~\cite{Hasenfratz:1998ri}. Hence, the overlap Dirac
operator provides a natural definition of the topological charge in the reduced
model. In fact, explicit gauge configurations which have a non-trivial
topological charge have been given in ref.~\cite{Kiskis:2002gr}.

A similar problem to define the topological charge (or the index) may be posed
in the context of a matrix model for the type IIB
superstring~\cite{Ishibashi:1996xs} which is a zero-dimensional reduction of
the $N=1$ super Yang-Mills theory in ten dimensions. In this context, a use of
the overlap-Dirac operator or the overlap~\cite{Narayanan:1993wx}
has been proposed~\cite{Kitsunezaki:1997iu}. (More precisely, this is for a
compact version of the IIB matrix model. See also ref.~\cite{Tada:1999mm}.) In
a related context, a use of Ginsparg-Wilson relation has been
actively investigated recently~\cite{Aoki:2002fq,Iso:2002jc}.

In a paper~\cite{Kikukawa:2002ms}, two of us studied chiral anomalies arising
from a fundamental fermion in the ``naive''~\cite{Eguchi:1982nm} or the
quenched~\cite{Bhanot:1982sh} reduced model. There, it was pointed out that a
certain restriction of reduced gauge fields (the $\U(1)$ embedding) allows a
mapping of the problem to that of lattice gauge theory. By using available
techniques in the latter, we determined a general form of the topological
charge resulted by a use of the overlap-Dirac operator. Also, in chiral-gauge
reduced models, it was shown that a single fundamental fermion gives rise to an
obstruction for a smooth fermion integration measure, which is analogous to the
gauge anomaly in the original theory before reduction.

An important question postponed in ref.~\cite{Kikukawa:2002ms} is how chiral
anomalies (which was evaluated only for a fundamental fermion) change as a
function of a gauge-group representation of the fermion. In particular, we are
interested in if the anomaly cancellation in the original theory is realized in
the chiral-gauge reduced model. As a step toward this investigation, in this
paper we study the topological charge (or the axial anomaly) in the reduced
model for various gauge-group representations.

In Section~2, we present a general setting of our problem in the naive or
the quenched reduced model. Next, in Section~3, we recapitulate basic reasoning
and results of ref.~\cite{Kikukawa:2002ms} concerning the axial anomaly. In
Section~4, on the basis of a structure of the overlap-Dirac operator, we
present general properties of the axial anomaly which can be stated without
any approximation. In subsequent sections, we perform a large~$N$ calculation
of the axial anomaly. To illustrate the idea of this calculational scheme, we
first re-derive the result of ref.~\cite{Kikukawa:2002ms} within this
approximation. Then, in Section~6, this scheme is applied to general
gauge-group representations. Section~7 is devoted to the conclusion.

\section{Reduced model with the overlap-Dirac operator}

The fermion sector of the vector-like reduced model is defined by
\begin{equation}
   \langle{\cal O}\rangle
   =\int\rmd\psi\rmd\overline\psi\,{\cal O}\exp(-\overline\psi D\psi),
\label{twoxone}
\end{equation}
where ${\cal O}$ is an arbitrary operator containing fermion variables and
$D$ is a Dirac operator. In this paper, we use the overlap-Dirac
operator~\cite{Neuberger:1998fp} as $D$. The overlap-Dirac operator is given by
\begin{equation}
   D=1-A(A^\dagger A)^{-1/2},\qquad A=1-D_{\rm w},
\label{twoxtwo}
\end{equation}
where $D_{\rm w}$ is the Wilson-Dirac operator:\footnote{We choose parameters
in the Wilson-Dirac operator as $m_0=r=1$. The Greek indices, $\mu$, $\nu$,
\dots, runs over 1, 2, \dots, $d$, where $d$ denotes a dimensionality of the
system which is assumed to be even.}
\begin{equation}
   D_{\rm w}={1\over2}
   \left[\gamma_\mu(\nabla_\mu^*+\nabla_\mu)+\sum_\mu(\nabla_\mu^*-\nabla_\mu)
   \right].
\label{twoxthree}
\end{equation}
The covariant derivatives in this expression depend on the gauge-group
representation~$R$, to which the fermion belongs. For the fundamental
representation which will be denoted by~$F$, they read
\begin{equation}
   \nabla_\mu\psi=U_\mu\psi-\psi,\qquad\nabla_\mu^*\psi=\psi-U_\mu^\dagger\psi,
\label{twoxfour}
\end{equation}
where $U_\mu\in\U(N)$ or $\SU(N)$ is the reduced gauge field.

The above prescription for the gauge coupling correspond to the ``naive''
reduced model~\cite{Eguchi:1982nm}. In the case of the
quenched reduced model~\cite{Bhanot:1982sh}, the Dirac operator should be
defined with a momentum insertion by the factor~$e^{ip_\mu}$. Since in this
paper we will treat the gauge field as a non-dynamical background, this phase
factor can be absorbed into the reduced gauge field without loss of generality.
So we will omit this momentum factor in the following discussion.

A general representation of the gauge group $\SU(N)$ is represented by
$(\psi)_{i_1,\ldots,i_n}$, where each of indices $i_1$, \dots, $i_n$ transforms
as the fundamental representation and indices may have certain symmetric
properties, as represented by the Young tableau. For example, the adjoint
fermion is expressed by~$(\psi)_{i_1,i_2,\ldots,i_N}$ where last $N-1$ indices
are totally anti-symmetric. This form of representation, however, has a certain
limitation in applying our large $N$ calculation, because our large $N$
calculation is justified only when the number of indices is $O(N^0)$. In the
above case of the adjoint fermion, we may equally express it as
$(\psi)_{i_1;j_1}$ by contracting last $N-1$ indices with the invariant tensor
so that the large~$N$ calculation is applied. With this situation in mind, we
consider a general representation expressed by
\begin{equation}
   (\psi)_{i_1,\ldots,i_n;j_1,\ldots,j_m},
\label{twoxfive}
\end{equation}
where each of indices $i_1$, \dots $i_n$ transforms as the fundamental
representation and each of $j_1$, \dots $j_m$ transforms as the
anti-fundamental representation. Indices may have certain symmetric
properties. We denote the structure of eq.~(\ref{twoxfive}) as $(n,m)$,
irrespective of its symmetry with respect to indices. For a representation
expressed by eq.~(\ref{twoxfive}), the covariant derivatives are defined by
\begin{eqnarray}
   &&(\nabla_\mu\psi)_{i_1,\ldots,i_n;j_1,\ldots,j_m}
\nonumber\\
   &&=(U_\mu)_{i_1k_1}\cdots(U_\mu)_{i_nk_n}
   (\psi)_{k_1,\ldots,k_n;l_1,\ldots,l_m}
   (U_\mu^\dagger)_{l_1j_1}\cdots(U_\mu^\dagger)_{l_mj_m}
   -(\psi)_{i_1,\ldots,i_n;j_1,\ldots,j_m},
\label{twoxsix}
\end{eqnarray}
and
\begin{eqnarray}
   &&(\nabla_\mu^*\psi)_{i_1,\ldots,i_n;j_1,\ldots,j_m}
\nonumber\\
   &&=(\psi)_{i_1,\ldots,i_n;j_1,\ldots,j_m}
   -(U_\mu^\dagger)_{i_1k_1}\cdots(U_\mu^\dagger)_{i_nk_n}
   (\psi)_{k_1,\ldots,k_n;l_1,\ldots,l_m}
   (U_\mu)_{l_1j_1}\cdots(U_\mu)_{l_mj_m}.
\label{twoxseven}
\end{eqnarray}

Now for the overlap-Dirac operator to be well-defined, we require that the
gauge field is ``admissible''~\cite{Hernandez:1999et,Kiskis:2002gr}. For the
fundamental representation, the condition reads
\begin{equation}
   \|1-U_\mu U_\nu U_\mu^\dagger U_\nu^\dagger\|<\epsilon,
\label{twoxeight}
\end{equation}
where $\epsilon$ is a constant smaller than~$(2-\sqrt{2})/d(d-1)$. For a
fermion in a general representation, the plaquette variable in this condition
is replaced by the plaquette in the corresponding gauge representation. In this
paper, however, we always require the condition~(\ref{twoxeight}) for the
fundamental representation, because we are interested in how the axial anomaly
changes as a function of the gauge-group representation. Namely, we consider
the anomaly for various gauge-group representations (including the fundamental
representation) by taking the {\it same\/} gauge-field configuration as the
gauge field background.

The overlap-Dirac operator satisfies the Ginsparg-Wilson
relation~\cite{Ginsparg:1982bj}
\begin{equation}
   \gamma_{d+1}D+D\gamma_{d+1}=D\gamma_{d+1}D,
\label{twoxnine}
\end{equation}
which implies an exact symmetry of the fermion action~\cite{Luscher:1998pq}:
The action $\overline\psi D\psi$ is invariant under substitutions,
$\psi\to\psi+\delta\psi$
and~$\overline\psi\to\overline\psi+\delta\overline\psi$, where
\begin{equation}
   \delta\psi=i\hat\gamma_{d+1}\psi,\qquad
   \delta\overline\psi=i\overline\psi\gamma_{d+1},
\label{twoxten}
\end{equation}
and $\hat\gamma_{d+1}$ is the modified chiral matrix defined by
\begin{equation}
   \hat\gamma_{d+1}=\gamma_{d+1}(1-D),\qquad \hat\gamma_{d+1}^2=1.
\label{twoxeleven}
\end{equation}
The fermion integration measure on the other hand acquires a non-trivial
jacobian~$Q_R$ (here the subscript~$R$ signifies the representation of the
fermion):
\begin{equation}
   \langle\delta{\cal O}\rangle
   =2iQ_R\langle{\cal O}\rangle,\qquad
   Q_R={1\over2}\mathop{\bf Tr}(\gamma_{d+1}+\hat\gamma_{d+1}),
\label{twoxtwelve}
\end{equation}
where ${\bf Tr}$ denotes a trace over a complete set which spans a space
of~$\psi$, including the trace over spinor indices. This combination~$Q_R$ is
an integer~\cite{Narayanan:1993wx,Hasenfratz:1998ri} again as a consequence of
the Ginsparg-Wilson relation and the $\gamma_5$-hermiticity,
$D^\dagger=\gamma_{d+1}D\gamma_{d+1}$. Hence we may regard this jacobian~$Q_R$
as a topological charge in the reduced
model~\cite{Kiskis:2002gr,Kikukawa:2002ms}.

\section{$\U(1)$ embedding and admissible $\U(1)$ gauge
fields\protect\footnote{This section gives a brief sketch of the basic
reasoning of ref.~\cite{Kikukawa:2002ms}. For details, the readers are asked to
refer to ref.~\cite{Kikukawa:2002ms}.}}

In ref.~\cite{Kikukawa:2002ms}, a general expression of the topological
charge~$Q_F$ (here the subscript~$F$ signifies the fundamental representation)
was given under a certain restriction of reduced gauge fields. This restriction
is termed $\U(1)$ embedding and defined as follows. First we assume that $N$ is
$d$-th power of a certain integer~$L$, $N=L^d$. Then the gauge field is assumed
to have the form
\begin{equation}
   U_\mu=u_\mu\Gammait_\mu,
\label{threexone}
\end{equation}
where~$u_\mu$ is an $N\times N$ {\it diagonal\/} matrix and $\Gammait_\mu$ is
the ``shift matrix''
\begin{equation}
   \Gammait_\mu={\mathbf 1}_L\otimes
   \cdots\otimes{\mathbf 1}_L\otimes U\otimes{\mathbf 1}_L
   \otimes\cdots\otimes{\mathbf 1}_L.
\label{threextwo}
\end{equation}
In this expression, ${\mathbf 1}_L$ represents the $L\times L$ unit matrix and
the factor~$U$ appears in the $\mu$-th entry. The $L\times L$ unitary
matrix~$U$ is given by
\begin{equation}
   U=\pmatrix{0&1&&\cr
              & \ddots & \ddots &\cr
              &        & \ddots &1\cr
              1&&&0\cr}.
\label{threexthree}
\end{equation}
When substituting eq.~(\ref{threexone}) in eq.~(\ref{twoxfour}), one finds that
covariant derivatives take an identical form as covariant derivatives in the
{\it conventional\/} lattice gauge theory. The size of the corresponding
lattice~$\Gammait$ is~$L$,
$\Gammait=\{\,x\in{\mathbb Z}^d\mid0\leq x_\mu<L\,\}$, and a
site~$x=(x_1,\ldots,x_d)$ on the lattice and an index of the fundamental
representation~$1\leq i\leq L^d=N$ are identified by
\begin{equation}
   i(x)=1+x_d+Lx_{d-1}+\cdots+L^{d-1}x_1.
\label{threexfour}
\end{equation}
Then the shift matrix~$\Gammait_\mu$ realizes a shift on the lattice in the
$\mu$-th direction,
\begin{equation}
   (\Gammait_\mu f\Gammait_\mu^\dagger)_{i(x)i(x)}
   =f_{i(x+\hat\mu)i(x+\hat\mu)},
\label{threexfive}
\end{equation}
for an arbitrary {\it diagonal\/} matrix~$f$. The gauge fields on the
lattice is given by diagonal elements of the matrix~$u_\mu$. Since $u_\mu$ is
unitary, all diagonal elements are pure-phase. Hence, the gauge field in the
reduced model which has the particular form~(\ref{threexone}) is mapped to
$\U(1)$ lattice gauge fields. Moreover, the plaquette variable in the reduced
model is identical to the plaquette variable in the $\U(1)$ lattice theory:
\begin{equation}
   U_\mu U_\nu U_\mu^\dagger U_\nu^\dagger
   =u_\mu(\Gammait_\mu u_\nu\Gammait_\mu^\dagger)
   (\Gammait_\nu u_\mu^\dagger\Gammait_\nu^\dagger)u_\nu^\dagger,
\label{threexsix}
\end{equation}
which is a diagonal matrix and its diagonal element is the $\U(1)$ plaquette
on~$\Gammait$:
\begin{eqnarray}
   &&(U_\mu U_\nu U_\mu^\dagger U_\nu^\dagger)_{i(x)i(x)}
\nonumber\\
   &&=(u_\mu)_{i(x)i(x)}(u_\nu)_{i(x+\hat\mu)i(x+\hat\mu)}
   (u_\mu)_{i(x+\hat\nu)i(x+\hat\nu)}^*(u_\nu)_{i(x)i(x)}^*
\nonumber\\
   &&=u_\mu(x)u_\nu(x+\hat\mu)u_\mu(x+\hat\nu)^*u_\nu(x)^*
\label{threexseven}
\end{eqnarray}
due to eq.~(\ref{threexfive}). In Appendix~A, we show that the above
restriction~(\ref{threexone}) is nothing but the
orbifolding~\cite{Douglas:1996sw}.

With the $\U(1)$ embedding, we can thus utilize available techniques in the
conventional $\U(1)$ lattice gauge theory. In particular, the above equivalence
of plaquette variables shows that the condition for admissible gauge
fields~(\ref{twoxeight}) is common in both matrix and lattice pictures. A
complete parameterization of admissible $\U(1)$ lattice gauge fields has been
known~\cite{Luscher:1999du}. In terms of the lattice gauge theory
on~$\Gammait$, it is given by
\begin{equation}
   u_\mu(x)=\omega(x)v_\mu^{[m]}(x)u_\mu^{[w]}(x)
   e^{ia_\mu^{\rm T}(x)}\omega(x+\hat\mu)^*.
\label{threexeight}
\end{equation}
In this expression, each factor has the following meaning: The field
$\omega(x)\in\U(1)$ is the $\U(1)$ gauge transformation (note that the
admissibility~(\ref{twoxeight}) is a gauge invariant condition). The
field~$u_\mu^{[w]}(x)$ is defined by
\begin{equation}
   u_\mu^{[w]}(x)=\cases{w_\mu,&for $x_\mu=0$,\cr
                         1,&otherwise,\cr}\qquad w_\mu\in\U(1),
\label{threexnine}
\end{equation}
and has a vanishing field strength~$f_{\mu\nu}(x)=0$, where
\begin{equation}
   f_{\mu\nu}(x)={1\over i}\ln
   u_\mu(x)u_\nu(x+\hat\mu)u_\mu(x+\hat\nu)^*u_\nu(x)^*,
   \qquad-\pi<f_{\mu\nu}(x)\leq\pi.
\label{threexten}
\end{equation}
However, the field~$u_\mu^{[w]}(x)$ carries the Wilson (or Polyakov) line,
$\prod_{s=0}^{L-1}u_\mu^{[w]}(s\hat\mu)=w_\mu\in\U(1)$. The
field~$v_\mu^{[m]}(x)$ is defined by
\begin{equation}
   v_\mu^{[m]}(x)=\exp\biggl[
   -{2\pi i\over L^2}\biggl(
   L\delta_{x_\mu,L-1}\sum_{\nu>\mu}m_{\mu\nu}x_\nu
   +\sum_{\nu<\mu}m_{\mu\nu}x_\nu
   \biggr)\biggr],
\label{threexeleven}
\end{equation}
and carries a constant field strength
\begin{equation}
   f_{\mu\nu}(x)={2\pi\over L^2}m_{\mu\nu},
\label{threextwelve}
\end{equation}
where the ``magnetic flux'' $m_{\mu\nu}$ is an {\it integer\/} bounded
by\footnote{$\epsilon'=2\arcsin(\epsilon/2)$.}
\begin{equation}
   |m_{\mu\nu}|<{\epsilon'\over2\pi}L^2.
\label{threexthirteen}
\end{equation}
The ``transverse'' gauge potential~$a_\mu^{\rm T}(x)$ is defined
by\footnote{$\partial_\mu$ and~$\partial_\mu^*$ denote the forward and the
backward difference operators on~$\Gammait$,
$\partial_\mu f(x)=f(x+\hat\mu)-f(x)$,
$\partial_\mu^*f(x)=f(x)-f(x-\hat\mu)$, respectively.}
\begin{eqnarray}
   &&\partial_\mu^*a_\mu^{\rm T}(x)=0,\qquad
   \sum_{x\in{\Gammait}}a_\mu^{\rm T}(x)=0,
\nonumber\\
   &&|f_{\mu\nu}(x)|
   =|\partial_\mu a_\nu^{\rm T}(x)-\partial_\nu a_\mu^{\rm T}(x)
   +2\pi m_{\mu\nu}/L^2|<\epsilon'.
\label{threexfourteen}
\end{eqnarray}
Note that the space of~$a_\mu^{\rm T}(x)$ is contractible. Namely, we may
smoothly deform as~$a_\mu^{\rm T}(x)\to0$ without being against the
admissibility.

The above is the $\U(1)$ embedding for the $\U(N)$ reduced model in which
$u_\mu$ in eq.~(\ref{threexone}) has no restriction except that it must be a
diagonal matrix. When the gauge group is~$\SU(N)$,
$\prod_{x\in{\mit\Gamma}}u_\mu(x)$ must be unity. This requires that
$w_\mu\in{\mathbb Z}_{L^{d-1}}$ and~$\prod_{x\in{\mit\Gamma}}v_\mu^{[m]}(x)=%
\exp[-\pi iL^{d-2}(L-1)\sum_\nu m_{\mu\nu}]=1$. Note that the latter is always
satisfied for~$d>2$.

When a local Dirac operator which obeys the Ginsparg-Wilson relation is
employed, we can moreover invoke a powerful cohomological technique which
determines a general structure of the axial anomaly in $\U(1)$ lattice gauge
theory~\cite{Luscher:1999kn}. Under the $\U(1)$ embedding, we can apply
this result in the lattice theory to the reduced model. In this way, we have
for the fundamental representation~\cite{Kikukawa:2002ms}
\begin{equation}
   Q_F={(-1)^{d/2}\over2^{d/2}(d/2)!}
   \epsilon_{\mu_1\nu_1\cdots\mu_{d/2}\nu_{d/2}}
   m_{\mu_1\nu_1}m_{\mu_2\nu_2}\cdots m_{\mu_{d/2}\nu_{d/2}},
\label{threexfifteen}
\end{equation}
which is manifestly an integer. Within the $\U(1)$ embedding, this is the
general form of~$Q_F$ defined by the overlap-Dirac operator.

Eq.~(\ref{threexfifteen}) establishes that the space of reduced gauge fields,
after the admissibility~(\ref{twoxeight}) is imposed, is divided into many
topological sectors. In ref.~\cite{Kiskis:2002gr}, this fact has been shown by
using the abelian background of ref.~\cite{Giusti:2001ta}. We emphasize that
this topological structure is {\it not\/} due to our restriction of reduced
gauge fields, the $\U(1)$ embedding~(\ref{threexone}). We
obtained~$Q_F$~(\ref{threexfifteen}) for a subspace of the whole space of
admissible reduced gauge fields, which can be expressed by
eq.~(\ref{threexone}). This is enough to conclude a non-trivial topological
structure of the space of admissible fields, because two configurations
(which are expressed by eq.~(\ref{threexone})) with different
$Q_F$~(\ref{threexfifteen}) cannot smoothly be connected even by relaxing the
constraint~(\ref{threexone}), without affecting the
admissibility~(\ref{twoxeight}). Recall that $Q_F$ itself is defined for all
admissible fields by eq.~(\ref{twoxtwelve}). The sole reason we stick to the
$\U(1)$ embedding is that it allows a precise parameterization of admissible
gauge fields~(\ref{threexeight}).

\section{General properties of~$Q_R$}

For the fundamental representation, a trick of $\U(1)$ embedding works
perfectly. If one is interested in the axial anomaly for other representations
with the {\it same\/} gauge-field background, however, the $\U(1)$ embedding
does not provide a useful picture such as the matrix-lattice correspondence.
We have to find another kind of argument. In this section, we summarize
general properties of~$Q_R$ which can be stated without any approximation.
We also present a simple argument within the $\U(1)$ embedding that
$Q_R=0$ for a representation~$R$ with $n-m=0\bmod L^2$ in eq.~(\ref{twoxfive}).

We start with the expression
\begin{eqnarray}
   &&Q_R={1\over2}\mathop{\bf Tr}{H\over\sqrt{H^2}},
\label{fourxone}\\
   &&H=\gamma_{d+1}A
   =\gamma_{d+1}\left[
   -{1\over2}\gamma_\mu(\nabla_\mu^*+\nabla_\mu)
   +1-{1\over2}\sum_\mu(\nabla_\mu^*-\nabla_\mu)\right],
\label{fourxtwo}
\end{eqnarray}
which follows from eqs.~(\ref{twoxtwelve}), (\ref{twoxeleven})
and~(\ref{twoxtwo}). Since $Q_R$ is given by a single trace over group indices,
for a direct-sum representation $R_1\oplus R_2$, we have
\begin{equation}
   Q_{R_1\oplus R_2}=Q_{R_1}+Q_{R_2}.
\label{fourxthree}
\end{equation}
Also for a complex conjugate representation~$R^*$, we have\footnote{This
follows from the property of the charge conjugation matrix~$C$ in
$d$-dimensional Euclidean space. It commutes with $\gamma_{d+1}$ in~$d=4k$
dimensions, while it anti-commutes with $\gamma_{d+1}$ in~$d=4k+2$ dimensions:
\[
   [C,\gamma_{d+1}]=0\quad(d=4k),\qquad
   \{C,\gamma_{d+1}\}=0\quad(d=4k+2).
\]
See ref.~\cite{Alvarez-Gaume:1985ex} for example.}
\begin{equation}
   Q_{R^*}=(-1)^{d/2}Q_R.
\label{fourxfour}
\end{equation}
These are expected algebraic properties as the axial anomaly.

Now, we assume the $\U(1)$ embedding~(\ref{threexone}). By noting
$\gamma_\mu\gamma_\nu=\delta_{\mu\nu}+[\gamma_\mu,\gamma_\nu]/2$, we have
\begin{eqnarray}
   H^2&=&-{1\over4}(\nabla_\mu^*+\nabla_\mu)^2
   +\left[1-{1\over2}\sum_\mu(\nabla_\mu^*-\nabla_\mu)\right]^2
\nonumber\\
   &&-{1\over8}\gamma_\mu\gamma_\nu
   [\nabla_\mu^*+\nabla_\mu,\nabla_\nu^*+\nabla_\nu]
   -{1\over4}\gamma_\mu\sum_\nu[\nabla_\mu^*+\nabla_\mu,
   \nabla_\nu^*-\nabla_\nu].
\label{fourxfive}
\end{eqnarray}
This shows that, in~$Q_R$~(\ref{fourxone}), the gamma matrices in the
denominator $\sqrt{H^2}$ always accompany a commutator of covariant
derivatives. A commutator of covariant derivatives can be computed from the
definition~(\ref{twoxsix}). The result is
\begin{eqnarray}
   &&([\nabla_\mu,\nabla_\nu]\psi)_{i_1,\ldots,i_n;j_1,\ldots,j_m}
\nonumber\\
   &&=(U_\mu U_\nu)_{i_1k_1}\cdots(U_\mu U_\nu)_{i_nk_n}
   (\psi)_{k_1,\ldots,k_n;l_1,\ldots,l_m}
   (U_\nu^\dagger U_\mu^\dagger)_{l_1j_1}\cdots
   (U_\nu^\dagger U_\mu^\dagger)_{l_mj_m}
\nonumber\\
   &&\quad-(U_\nu U_\mu)_{i_1k_1}\cdots(U_\nu U_\mu)_{i_nk_n}
   (\psi)_{k_1,\ldots,k_n;l_1,\ldots,l_m}
   (U_\mu^\dagger U_\nu^\dagger)_{l_1j_1}\cdots
   (U_\mu^\dagger U_\nu^\dagger)_{l_mj_m}.
\label{fourxsix}
\end{eqnarray}
This involves an exchange~$U_\mu\leftrightarrow U_\nu$ which can be expressed
by the plaquette variable as
\begin{equation}
   U_\mu U_\nu=(U_\mu U_\nu U_\mu^\dagger U_\nu^\dagger)U_\nu U_\mu
   =e^{if_{\mu\nu}}U_\nu U_\mu,
\label{fourxseven}
\end{equation}
where we have used the relation~(\ref{threexseven}) for the $\U(1)$ embedding
and here $f_{\mu\nu}$ is a diagonal matrix whose diagonal elements are given by
eq.~(\ref{threexten}) for the parameterization~(\ref{threexeight}).

Since $Q_R$ is an integer, it is invariant under a deformation of the gauge
field, as long as the deformation is not against the
admissibility~(\ref{twoxeight}). In particular, we may remove the degrees of
freedom $a_\mu^{\rm T}(x)$ in eq.~(\ref{threexeight}), because a deformation
$a_\mu^{\rm T}(x)\to0$ does not affect the admissibility. Then the field
strength of admissible gauge fields is given by the
{\it constant\/}~(\ref{threextwelve}). Namely, we can set
\begin{equation}
   U_\mu U_\nu=e^{2\pi im_{\mu\nu}/L^2}U_\nu U_\mu,
\label{fourxeight}
\end{equation}
in evaluation of~$Q_R$. This shows
\begin{eqnarray}
   &&([\nabla_\mu,\nabla_\nu]\psi)_{i_1,\ldots,i_n;j_1,\ldots,j_m}
\nonumber\\
   &&=\left[e^{2\pi i(n-m)m_{\mu\nu}/L^2}-1\right]
\nonumber\\
   &&\quad\times
   (U_\nu U_\mu)_{i_1k_1}\cdots(U_\nu U_\mu)_{i_nk_n}
   (\psi)_{k_1,\ldots,k_n;l_1,\ldots,l_m}
   (U_\mu^\dagger U_\nu^\dagger)_{l_1j_1}\cdots
   (U_\mu^\dagger U_\nu^\dagger)_{l_mj_m},
\label{fourxnine}
\end{eqnarray}
and similar considerations show
\begin{eqnarray}
   &&([\nabla_\mu,\nabla_\nu^*]\psi)_{i_1,\ldots,i_n;j_1,\ldots,j_m}
\nonumber\\
   &&=\left[1-e^{-2\pi i(n-m)m_{\mu\nu}/L^2}\right]
\nonumber\\
   &&\quad\times
   (U_\nu^\dagger U_\mu)_{i_1k_1}\cdots(U_\nu^\dagger U_\mu)_{i_nk_n}
   (\psi)_{k_1,\ldots,k_n;l_1,\ldots,l_m}
   (U_\mu^\dagger U_\nu)_{l_1j_1}\cdots
   (U_\mu^\dagger U_\nu)_{l_mj_m},
\label{fourxten}
\end{eqnarray}
and
\begin{eqnarray}
   &&([\nabla_\mu^*,\nabla_\nu^*]\psi)_{i_1,\ldots,i_n;j_1,\ldots,j_m}
\nonumber\\
   &&=\left[e^{2\pi i(n-m)m_{\mu\nu}/L^2}-1\right]
\nonumber\\
   &&\quad\times
   (U_\nu^\dagger U_\mu^\dagger)_{i_1k_1}
   \cdots(U_\nu^\dagger U_\mu^\dagger)_{i_nk_n}
   (\psi)_{k_1,\ldots,k_n;l_1,\ldots,l_m}
   (U_\mu U_\nu)_{l_1j_1}\cdots
   (U_\mu U_\nu)_{l_mj_m}.
\label{fourxeleven}
\end{eqnarray}
As a consequence, we see that commutators of covariant derivatives vanish
when~$n-m=0\bmod L^2$.

As a side remark, we note that the factor $e^{2\pi i(n-m)m_{\mu\nu}/L^2}$
appearing in above expressions is the plaquette variable in the
representation~$R$.\footnote{Hence these expressions show that the
configuration~(\ref{threexeight}) with $a_\mu^{\rm T}(x)=0$ in fact satisfies
the admissibility condition for a representation~$R$, if
$|[(n-m)\bmod L^2]m_{\mu\nu}|<\epsilon'L^2/(2\pi)$.} This factor is simply
proportional to the identity matrix and when the gauge group is~$\SU(N)$, this
implies that the plaquette has its value in the center of $\SU(N)$. When $L$ is
large, the plaquette behaves as $1+2\pi i(n-m)m_{\mu\nu}/L^2$ which does
{\it not\/} have a structure as one plus (traceless) Lie algebra valued matrix,
although the plaquette itself belongs to $\SU(N)$. An implication of this fact
will be commented later.

As we have noted, if commutators of covariant derivatives vanish, we have no
gamma matrices in the denominator of eq.~(\ref{fourxone}), $\sqrt{H^2}$. Then
we cannot have an enough number of gamma matrices to have non-zero trace with
respect to spinor indices in~$Q_R$. Therefore we conclude
\begin{equation}
   Q_R=0,\qquad{\rm for}\qquad n-m=0\pmod{L^2}.
\label{fourxtwelve}
\end{equation}
For an irreducible representation of~$\SU(N)$ expressed by the Young tableau,
the condition $n-m=0\bmod{L^2}$ implies that the total number of boxes is a
multiple of~$L^2$. In particular, since the adjoint representation corresponds
to $(n,m)=(1,1)$ (or equivalently $(n,m)=(N,0)$), $Q_R=0$ for the adjoint
fermion within the $\U(1)$ embedding. A tensor-product representation of such
irreducible representations also has $n-m=0\bmod{L^2}$. So it also leads
$Q_R=0$.

\section{$Q_F$ in the large~$N$ limit}

It seems difficult to conclude something exact about~$Q_R$ for general
gauge-group representations than eq.~(\ref{fourxtwelve}). In what follows, we
will consider an approximate treatment in the large $N$ or equivalently the
large~$L$ limit. Our scheme was inspired by a calculation of
ref.~\cite{Iso:2002jc}. Although our target is general gauge-group
representations, it is helpful to consider first the fundamental
representation in this limit to illustrate the idea.

First to give a precise meaning of the trace in eq.~(\ref{fourxone}), we have
to introduce some complete basis for the fermion variable. For this, we use
eigenvectors of the shift matrix
\begin{equation}
   \Gammait_\mu\phi(\vec p)
   =e^{2\pi ip_\mu/L}\phi(\vec p),\qquad
   \{\,\vec p\in{\mathbb Z}^d\mid0\leq p_\mu<L\,\}.
\label{fivexone}
\end{equation}
The explicit form of $\phi(\vec p)$ is given by
\begin{equation}
   \phi(\vec p)={1\over\sqrt{N}}\hat\phi(\vec p)
   \pmatrix{1\cr1\cr\vdots\cr1\cr},
\label{fivextwo}
\end{equation}
where the $N\times N$ matrix $\hat\phi(\vec p)$ is defined by
\begin{equation}
   \hat\phi(\vec p)=V^{p_1}\otimes\cdots\otimes V^{p_d},\qquad
   V=\pmatrix{1& & & & \cr
                 &\omega &&&\cr
                 &       &\omega^2&&\cr
                 &       &        &\ddots&\cr
                 &       &        &      &\omega^{L-1}\cr},
\label{fivexthree}
\end{equation}
with $\omega=e^{2\pi i/L}$. One can easily verify eq.~(\ref{fivexone}) and
\begin{equation}
   \phi(\vec p)^\dagger\phi(\vec q)
   =\delta_{\vec p,\vec q}.
\label{fivexfour}
\end{equation}
It can also be shown that $\phi(\vec p)$ span a complete basis (Appendix~B). So
we have
\begin{equation}
   \sum_{\vec p}\phi(\vec p)\phi(\vec p)^\dagger={\mathbf 1}_N.
\label{fivexfive}
\end{equation}
A fermion in the fundamental representation can then be expanded in this basis
as
\begin{equation}
   \psi_i=\sum_{\vec p}\phi(\vec p)_ic(\vec p).
\label{fivexsix}
\end{equation}
The topological charge~(\ref{fourxone}) for the fundamental fermion is then
given by
\begin{equation}
   Q_F
   ={1\over2}\sum_{\vec p}\phi(\vec p)^\dagger{H\over\sqrt{H^2}}\phi(\vec p).
\label{fivexseven}
\end{equation}

Now, our large $N$ calculation of~$Q_F$ proceeds as follows. First, for
configurations in the $\U(1)$ embedding~(\ref{threexone}) with
eq.~(\ref{threexeight}), we set $\omega(x)=1$ and~$a_\mu^{\rm T}(x)=0$. These
choices do not affect $Q_F$ because $Q_F$ is gauge invariant and the integer
$Q_F$ does not change under the deformation $a_\mu^{\rm T}(x)\to0$ that is
consistent with the admissibility. Next, we assume that in
eq.~(\ref{threexeight}),
\begin{equation}
   u_\mu={\mathbf 1}_N(1+O(1/L)),
\label{fivexeight}
\end{equation}
or equivalently
\begin{equation}
   U_\mu=\Gammait_\mu(1+O(1/L)).
\label{fivexnine}
\end{equation}
{}From the explicit form of admissible gauge fields, we see that the assumption
is fulfilled except for the factor~$u_\mu^{[w]}(x)$ (the Wilson-line degrees of
freedom) of eq.~(\ref{threexnine}). We note however that for a fixed value of
the Wilson line~$w_\mu$, one may choose a different representation
of~$u_\mu^{[w]}(x)$ by making use of the gauge degrees of freedom. A possible
choice is
\begin{equation}
   u_\mu^{[w]}(x)=(w_\mu)^{1/L},
\label{fivexten}
\end{equation}
which in fact reproduces an identical Wilson line
$\prod_{s=0}^{L-1}u_\mu^{[w]}(s\hat\mu)=w_\mu$. Then the
condition~(\ref{fivexeight}) is fulfilled for eq.~(\ref{fivexten}). Once the
condition~(\ref{fivexnine}) matches, it is straightforward to find the large
$N$ limit (or the large $L$ limit) of~$Q_F$.

Recall the structure of~$H^2$~(\ref{fourxfive}). Noting that
eqs.~(\ref{fourxnine})--(\ref{fourxeleven}) show that commutators of covariant
derivatives are proportional to $2\pi im_{\mu\nu}/L^2$ in the large~$L$ limit,
we see that the leading term in the large~$N$ limit is contained in an
expansion of
\begin{eqnarray}
   Q_F&=&{1\over2}\sum_{\vec p}\tr\gamma_{d+1}
   \left[-i\gamma_\mu s_\mu+1+\sum_\mu(c_\mu-1)\right]
\nonumber\\
   &&\times\left\{
   s_\nu^2+\left[1+\sum_\nu(c_\nu-1)\right]^2
   -{2\pi i m_{\nu\rho}\over 2L^2}\gamma_\nu\gamma_\rho c_\nu c_\rho
   +{2\pi im_{\nu\rho}\over L^2}\gamma_\nu ic_\nu s_\rho\right\}^{-1/2},
\label{fivexeleven}
\end{eqnarray}
where we have noted eqs.~(\ref{fivexnine}) and~(\ref{fivexone}) and used
abbreviations
\begin{equation}
   s_\mu=\sin{2\pi\over L}p_\mu,\qquad c_\mu=\cos{2\pi\over L}p_\mu.
\label{fivextwelve}
\end{equation}
The expression~(\ref{fivexeleven}) is familiar in the {\it classical continuum
limit\/} of the axial anomaly which is defined by the overlap-Dirac operator in
the conventional lattice gauge theory~\cite{Kikukawa:1998pd}. See eq.~(12) of
ref.~\cite{Fujiwara:2002xh}. The lattice spacing~$a$ in the latter is formally
identified with~$1/L$ in our expression. The correspondence becomes perfect in
the large $N$ limit, in which
\begin{equation}
   k_\mu={2\pi\over L}p_\mu,
\label{fivexthirteen}
\end{equation}
can be regarded as a continuous momentum and thus the sum over $\vec p$
becomes momentum integration
\begin{equation}
   \sum_{\vec p}\to L^d\int_{-\pi}^\pi{\rmd^dk\over(2\pi)^d}.
\label{fivexfourteen}
\end{equation}
Therefore we can literally copy the result of ref.~\cite{Fujiwara:2002xh}. The
large $N$ limit of~$Q_F$ is then given by
\begin{equation}
   Q_F={(-1)^{d/2}\over2^{d/2}(d/2)!}
   \epsilon_{\mu_1\nu_1\cdots\mu_{d/2}\nu_{d/2}}
   m_{\mu_1\nu_1}m_{\mu_2\nu_2}\cdots m_{\mu_{d/2}\nu_{d/2}},
\label{fivexfifteen}
\end{equation}
which coincides with eq.~(\ref{threexfifteen}). Eq.~(\ref{threexfifteen}) was
originally obtained for finite~$N$ by using the local cohomology
technique and here we computed it in the large $N$ limit. However, since
$Q_F$~(\ref{threexfifteen}) is independent of~$N$, a coincide with the
large~$N$ limit~(\ref{fivexfifteen}) is expected. Nevertheless, this
coincidence is assuring: In fact the above large $N$ calculation might be
regarded as another proof of eq.~(\ref{threexfifteen}).

\section{$Q_R$ in the large~$N$ limit}

An advantage of the large $N$ approach is that it is applicable to higher
representations for which an application of the local cohomology technique is
difficult.\footnote{By using techniques in refs.~\cite{Connes:1997cr}, it is
possible to map the fermion sector of the reduced model to a noncommutative
field theory. However, with a general lack of locality in noncommutative field
theories, it seems rather difficult to develop a local cohomological argument
in these theories.} First, it is straightforward to generalize the above large
$N$ calculation for the fundamental representation to a tensor-product of
(anti-)fundamental representations. A fermion in the tensor-product
representation can be expanded in an analogous way as eq.~(\ref{fivexsix})
\begin{eqnarray}
   (\psi)_{i_1,\ldots,i_n;j_1,\ldots,j_m}
   &=&\sum_{\vec p_1}\cdots\sum_{\vec p_n}
   \sum_{\vec q_1}\cdots\sum_{\vec q_m}
   \phi(\vec p_1)_{i_1}\cdots\phi(\vec p_n)_{i_n}
   \phi(\vec q_1)_{j_1}^\dagger\cdots\phi(\vec q_m)_{j_m}^\dagger
\nonumber\\
   &&\times c(\vec p_1,\ldots,\vec p_n;\vec q_1,\ldots,\vec q_m).
\label{sixxone}
\end{eqnarray}
The topological charge~(\ref{fourxone}) for a $(n,m)$-type tensor-product is
then given by
\begin{eqnarray}
   Q_{(n,m)}&=&{1\over2}
   \sum_{\vec p_1}\cdots\sum_{\vec p_n}
   \sum_{\vec q_1}\cdots\sum_{\vec q_m}
   \phi(\vec p_1)_{i_1}^\dagger\cdots\phi(\vec p_n)_{i_n}^\dagger
   \phi(\vec q_1)_{j_1}\cdots\phi(\vec q_m)_{j_m}
\nonumber\\
   &&\qquad
   \times\left({H\over\sqrt{H^2}}\right)_{i_1,\ldots,i_n;j_1,\ldots,j_m;
   k_1,\ldots,k_n;l_1,\ldots,l_m}
\nonumber\\
   &&\qquad\qquad\times\phi(\vec p_1)_{k_1}\cdots\phi(\vec p_n)_{k_n}
   \phi(\vec q_1)_{l_1}^\dagger\cdots\phi(\vec q_m)_{l_m}^\dagger.
\label{sixxtwo}
\end{eqnarray}
The following steps are almost identical as for the fundamental
representation. Noting eqs.~(\ref{fourxnine})--(\ref{fourxeleven}), we find the
expression~(\ref{fivexeleven}) with the following substitution:
\begin{equation}
   \sum_{\vec p}\to\sum_{\vec p_1}\cdots\sum_{\vec p_n}
   \sum_{\vec q_1}\cdots\sum_{\vec q_m},
\label{sixxthree}
\end{equation}
and
\begin{equation}
   m_{\mu\nu}\to(n-m)m_{\mu\nu},
\label{sixxfour}
\end{equation}
and
\begin{eqnarray}
   &&s_\mu\to\sin{2\pi\over L}
   (\vec p_1+\cdots+\vec p_n-\vec q_1-\cdots-\vec q_m)_\mu,
\nonumber\\
   &&c_\mu\to\cos{2\pi\over L}
   (\vec p_1+\cdots+\vec p_n-\vec q_1-\cdots-\vec q_m)_\mu.
\label{sixxfive}
\end{eqnarray}
Here we have assumed that $n$ and $m$ are of~$O(N^0)$, i.e., they are not large
numbers. This assumption eliminates a possibility that a large number of link
variables in a covariant derivative compensate a ``weakness'' of gauge fields
in the large $N$ limit~(\ref{fivexnine}). Then the argument for the fundamental
representation with above substitutions proceeds as it stands.

In eq.~(\ref{fivexeleven}) with above substitutions, we shift the variable, say
$\vec p_1$,
as $\vec p_1\to \vec p_1-\vec p_2\cdots-\vec p_n+\vec q_1+\cdots+\vec q_m$
so that the ``integrand'' becomes independent of $\vec p_2$, \dots, $\vec p_n$,
$\vec q_1$, \dots, $\vec q_m$. The summation over these variables can then be
done and it gives a factor
\begin{equation}
   \sum_{\vec p_2}\cdots\sum_{\vec p_n}
   \sum_{\vec q_1}\cdots\sum_{\vec q_m}1=(L^d)^{n+m-1}=N^{n+m-1}.
\label{sixxsix}
\end{equation}
The summation over $\vec p_1$ becomes the integral~(\ref{fivexfourteen}) in the
large $N$ limit. In this way, we have
\begin{equation}
   Q_{(n,m)}=(n-m)^{d/2}N^{n+m-1}Q_F(1+O(1/L)),
   \qquad{\rm for}\qquad n,m=O(N^0).
\label{sixxseven}
\end{equation}
Note that $N^{n+m-1}=(\dim R)/N$ for the tensor-product representation~$R$.

With the above experience, let us consider the large $N$ calculation for higher
irreducible representations. For general irreducible representation~$R$, the
topological charge~(\ref{fourxone}) can be expressed as
\begin{eqnarray}
   Q_{(n,m)}&=&{1\over2}
   \sum_{\vec p_1}\cdots\sum_{\vec p_n}
   \sum_{\vec q_1}\cdots\sum_{\vec q_m}
   \phi(\vec p_1)_{i_1}^\dagger\cdots\phi(\vec p_n)_{i_n}^\dagger
   \phi(\vec q_1)_{j_1}\cdots\phi(\vec q_m)_{j_m}
\nonumber\\
   &&\qquad
   \times\left({H\over\sqrt{H^2}}{\cal P}_R\right)_{i_1,\ldots,i_n;j_1,
   \ldots,j_m;k_1,\ldots,k_n;l_1,\ldots,l_m}
\nonumber\\
   &&\qquad\qquad\times\phi(\vec p_1)_{k_1}\cdots\phi(\vec p_n)_{k_n}
   \phi(\vec q_1)_{l_1}^\dagger\cdots\phi(\vec q_m)_{l_m}^\dagger,
\label{sixxeight}
\end{eqnarray}
where ${\cal P}_R$ is the projection operator for the irreducible
representation~$R$. For the anti-symmetric representation, for example, it
reads
\begin{equation}
   \left({\cal P}_R\right)_{i_1i_2;;k_1k_2;}
   ={1\over2}\left(\delta_{i_1k_1}\delta_{i_2k_2}
   -\delta_{i_1k_2}\delta_{i_2k_1}\right).
\label{sixxnine}
\end{equation}
For other irreducible representations (that can be obtained by appropriate
(anti-) symmetrization of indices of a tensor-product representation), it is
straightforward to construct~${\cal P}_R$. In our large~$N$ calculation, we do
not need the explicit form of~${\cal P}_R$; what we need is the weight~$w$ of
the ``identity operator'' in ${\cal P}_R$:
\begin{eqnarray}
   &&\left({\cal P}_R\right)_{i_1,\ldots,i_n;j_1,
   \ldots,j_m;k_1,\ldots,k_n;l_1,\ldots,l_m}
\nonumber\\
   &&
   =w\delta_{i_1k_1}\cdots\delta_{i_nk_n}\delta_{j_1l_1}\cdots\delta_{j_ml_m}
   +\left(\widetilde{\cal P}_R\right)_{i_1,\ldots,i_n;j_1,
   \ldots,j_m;k_1,\ldots,k_n;l_1,\ldots,l_m},
\label{sixxten}
\end{eqnarray}
where the operator~$\widetilde{\cal P}_R$ exchanges at least one index-pair.
To find a value of~$w$, we note the relation
\begin{eqnarray}
   &&\sum_{\vec p_1}\cdots\sum_{\vec p_n}
   \sum_{\vec q_1}\cdots\sum_{\vec q_m}
   \phi(\vec p_1)_{i_1}^\dagger\cdots\phi(\vec p_n)_{i_n}^\dagger
   \phi(\vec q_1)_{j_1}\cdots\phi(\vec q_m)_{j_m}
\nonumber\\
   &&\qquad
   \times\left({\cal P}_R\right)_{i_1,\ldots,i_n;j_1,
   \ldots,j_m;k_1,\ldots,k_n;l_1,\ldots,l_m}
\nonumber\\
   &&\qquad\qquad\times\phi(\vec p_1)_{k_1}\cdots\phi(\vec p_n)_{k_n}
   \phi(\vec q_1)_{l_1}^\dagger\cdots\phi(\vec q_m)_{l_m}^\dagger
\nonumber\\
      &&=\dim R
\nonumber\\
      &&=wN^d(1+O(1/N)),
\label{sixxeleven}
\end{eqnarray}
where $\dim R$ is the dimension of the representation~$R$. This relation holds
because the operator~$\widetilde{\cal P}_R$ produces a delta function of a pair
of momenta which restricts the ``phase-space'' integral by a factor~$N$.
Therefore, we have
\begin{equation}
   w={\dim R\over N^d}(1+O(1/N)).
\label{sixxtwelve}
\end{equation}

Now, when acting on $\phi(\vec p_1)_{k_1}\cdots\phi(\vec p_n)_{k_n}
\phi(\vec q_1)_{l_1}^\dagger\cdots\phi(\vec q_m)_{l_m}^\dagger$, the projection
operator~${\cal P}_R$ exchanges the name of ``wave functions'' within the
set~$\{\phi(\vec p_1),\ldots,\phi(\vec p_n)\}$ and
within $\{\phi(\vec q_1)^\dagger,\ldots,\phi(\vec q_m)^\dagger\}$, but not
extending these two sets. Since the action of $H/\sqrt{H^2}$ (in the large $N$
limit) is invariant under this exchange, as can be seen from
eq.~(\ref{sixxfive}), what is left after the same argument as before is
\begin{eqnarray}
   Q_F&=&{1\over2}\sum_{\vec p_1}\cdots\sum_{\vec p_n}
   \sum_{\vec q_1}\cdots\sum_{\vec q_m}
   \phi(\vec p_1)_{i_1}^\dagger\cdots\phi(\vec p_n)_{i_n}^\dagger
   \phi(\vec q_1)_{j_1}\cdots\phi(\vec q_m)_{j_m}
\nonumber\\
   &&\qquad\times\left({\cal P}_R\right)_{i_1,\ldots,i_n;j_1,
   \ldots,j_m;k_1,\ldots,k_n;l_1,\ldots,l_m}
   \phi(\vec p_1)_{k_1}\cdots\phi(\vec p_n)_{k_n}
   \phi(\vec q_1)_{l_1}^\dagger\cdots\phi(\vec q_m)_{l_m}^\dagger
\nonumber\\
   &&\qquad\times
   \tr\gamma_{d+1}
   \left[-i\gamma_\mu s_\mu+1+\sum_\mu(c_\mu-1)\right]
\nonumber\\
   &&\qquad\times\Biggl\{
   s_\nu^2+\left[1+\sum_\nu(c_\nu-1)\right]^2
\nonumber\\
   &&\qquad\qquad
   -{2\pi i (n-m)m_{\nu\rho}\over 2L^2}\gamma_\nu\gamma_\rho c_\nu c_\rho
   +{2\pi i(n-m)m_{\nu\rho}\over L^2}\gamma_\nu ic_\nu s_\rho
   \Biggr\}^{-1/2},
\label{sixxthirteen}
\end{eqnarray}
where $s_\mu$ and~$c_\mu$ are given by eq.~(\ref{sixxfive}). By substituting
eq.~(\ref{sixxten}), the operator~$\widetilde{\cal P}_R$ produces only a
sub-leading contribution of~$O(1/N)$. So, by only taking a contribution of the
identity operator, the integration over (say) $\vec p_1$ can be done in the
large $N$ limit as before. In this way, we finally have\footnote{An interesting
observation which we have verified for a wide class of gauge-group
representations is that the coefficient $(n-m)\dim R/N$ is an {\it integer}.
We have no proof of this statement at the moment. If this generally holds,
we would speculate $Q_R=(n-m\bmod L^2)^{d/2}(\dim R)Q_F/N$ is an exact
expression being valid even for finite~$N$.}
\begin{equation}
   Q_R=(n-m)^{d/2}{\dim R\over N}Q_F(1+O(1/L)),
   \qquad{\rm for}\qquad n,m=O(N^0).
\label{sixxfourteen}
\end{equation}
This large~$N$ result in fact reproduces our previous results;
eq.~(\ref{threexfifteen}) when $(n,m)=(1,0)$, eq.~(\ref{fourxtwelve})
when $n=m$ and eq.~(\ref{sixxseven}) for tensor-product representations. One
can express this anomaly in an analogous form to the Chern character: Introduce
a 2-form
\begin{equation}
   M_R={1\over2}(n-m)_R\,m_{\mu\nu}\,\rmd x_\mu\rmd x_\nu,
\label{sixxfifteen}
\end{equation}
then
\begin{equation}
   Q_R\,\rmd^dx={1\over N}\tr_R\exp(-M_R)\Bigr|_{\rmd^dx}(1+O(1/L)).
\end{equation}
We note algebraic properties of~$M_R$,
$M_{R_1\otimes R_2}=M_{R_1}+M_{R_2}$ and~$M_{R^*}=-M_R$ ($R\to R^*$ is
equivalent to an exchange~$n\leftrightarrow m$), which is analogous to that of
the field strength 2-form. Our $Q_R$ in the large $N$ limit thus satisfies
eq.~(\ref{fourxfour}). For a direct-product of two representations, we have
\begin{eqnarray}
   Q_{R_1\otimes R_2}\rmd^dx
   &=&{(-1)^{d/2}\over(d/2)!}{1\over N}\tr_{R_1\otimes R_2}
   (M_{R_1}+M_{R_2})^{d/2}
\nonumber\\
   &=&Q_{R_1}\dim R_2\,\rmd^dx+Q_{R_2}\dim R_1\,\rmd^dx+\cdots.
\end{eqnarray}
This is an expected algebraic property as the axial anomaly. Recall however
that with the $\U(1)$ embedding, the plaquette of the reduced gauge field has
its value in the center of~$\SU(N)$. So even when the gauge group is $\SU(N)$,
$\tr_R M_R\neq0$. The axial anomaly we found thus has a different structure
from the axial anomaly in the continuum $\SU(N)$ gauge field theory, in which
the anomaly is given by the trace of a product of traceless gauge-group
generators.

\section{Conclusion}

In this paper, we have studied the axial anomaly arising from the fermion
sector of the naive or the quenched reduced model. We used the overlap-Dirac
operator and consider a restriction of reduced gauge fields by $\U(1)$
embedding. Our main results are eq.~(\ref{fourxtwelve}) and
eq.~(\ref{sixxfourteen}) for the axial anomaly for general gauge-group
representations. We expect that our analyses in this paper will be useful to
investigate a possible gauge anomaly in the reduced
model~\cite{Kikukawa:2002ms} for general gauge-group representations and its
cancellation among fermion multiplet.

H.S. would like to thank Tohru Eguchi, Kazuo Fujikawa, Takanori Fujiwara,
Yusuke Taniguchi and Ke Wu for useful discussions. This work was completed when
we attended RIKEN topical meeting, ``Chiral fermion on the lattice and geometry
of matrix models''. We thank the organizer and participants, in particular,
Masashi Hayakawa, Satoshi Iso, Hikaru Kawai and Tsukasa Tada, for their warm
hospitality and discussions. This work is supported in part by Grant-in-Aid
for Scientific Research, \#12640262, \#14046207 (Y.K.) and \#13740142 (H.S.).

\appendix

\section{$\U(1)$ embedding is the orbifolding\protect\footnote{This fact was
suggested to us by Yusuke Taniguchi. A similar observation from a more general
view point has been presented in ref.~\cite{Neuberger:2002bk}. An observation
that a matrix model with appropriate constraints can be regarded as a lattice
theory dates back to the early study on the reduced
model~\cite{Okawa:1982ic}.}}

We define a diagonal matrix~$f$
\begin{equation}
   f=e^{-\pi iL(N-1)/N^2}
   \Lambdait^{L^{d-1}}\otimes\Lambdait^{L^{d-2}}\otimes\cdots
   \otimes\Lambdait^L\otimes\Lambdait,
\label{axone}
\end{equation}
where
\begin{equation}
   \Lambdait=\pmatrix{1& & & & \cr
                 &\lambda &&&\cr
                 &       &\lambda^2&&\cr
                 &       &        &\ddots&\cr
                 &       &        &      &\lambda^{L-1}\cr},\qquad
   \lambda=e^{2\pi i/N}=e^{2\pi i/L^d}.
\end{equation}
In eq.~(\ref{axone}), the factor~$e^{-\pi iL(N-1)/N^2}$ is multiplied so that
$\det f=1$. Namely, $f\in\SU(N)$. In the notation of eq.~(\ref{threexfour}),
the diagonal element of~$f$ is given by
\begin{equation}
   f_{i(x)i(x)}=e^{-\pi iL(N-1)/N^2}\lambda^{i(x)-1}.
\end{equation}
Since the integer $i(x)-1$ runs over all integers from~$0$ to~$N-1$, all
diagonal elements of~$f$ are distinct pure-phase. Also, one can verify
\begin{equation}
   \Gammait_\mu f\Gammait_\mu^\dagger=\lambda^{L^{d-\mu}}f.
\end{equation}

Now we impose a constraint on the reduced gauge field\footnote{Note however
that in the $\U(1)$ embedding we do not place any restriction on reduced
fermion fields.}
\begin{equation}
   U_\mu=\lambda^{L^{d-\mu}}fU_\mu f^\dagger.
\label{axfive}
\end{equation}
In terms of a matrix~$u_\mu$ defined by~$U_\mu=u_\mu\Gammait_\mu$, this reads
\begin{equation}
   u_\mu=fu_\mu f^\dagger.
\end{equation}
Since $f$ is a diagonal matrix whose all elements are distinct pure-phase,
this implies that $u_\mu$ is a diagonal matrix. Namely, the
constraint~(\ref{axfive}) is equivalent to the $\U(1)$
embedding~(\ref{threexone}). In eq.~(\ref{axfive}), the conjugation
by~$f\in\SU(N)$ can be regarded as the gauge transformation in the reduced
model and a multiplication of the factor~$\lambda^{L^{d-\mu}}\in\U(1)$ can be
regarded as~$\U(1)^d$ transformation, under which the pure-gauge reduced model
(i.e., the plaquette action) is
invariant~\cite{Eguchi:1982nm}--\cite{Das:1984nb}. In this sense, the
constraint~(\ref{axfive}) can be regarded as the orbifold
projection~\cite{Douglas:1996sw}. Namely, the $\U(1)$ embedding is a simple
example of the orbifolding. Among the $\U(N)$ or $\SU(N)$ gauge transformations
in the reduced model, $U_\mu\to\Omega U_\mu\Omega^\dagger$, those which are
consistent with the projection~(\ref{axfive}) survive as the gauge symmetry in
the $\U(1)$ embedding. These are given by diagonal~$\Omega$'s and are nothing
but $\U(1)$ gauge transformations in the $\U(1)$ embedding, discussed
in~ref.~\cite{Kikukawa:2002ms}.

\section{$\phi(\vec p)$ spans a complete basis}

Let $y$ be an integer, $y=1$, $2$, \dots, $L-1$. Then $(\omega^y)^L=1$ but
$\omega^y\neq1$. Defining a polynomial $P(M)=\sum_{p=0}^{L-1}M^p/L$, these
imply
\begin{equation}
   P(\omega^y)=0,\qquad P(1)=1.
\end{equation}
Then an inspection of the structure of~$V$ shows
\begin{equation}
   P(V\omega^{-x})=\mathop{\rm diag}(0,\ldots,0,1,0,\ldots,0),\qquad
   x=0,1,\ldots,L-1,
\end{equation}
where the non-zero element appears in the $x+1$-th entry. Then we see that
the $N$-vector
\begin{eqnarray}
   P(V\omega^{-x_1})\otimes\cdots\otimes P(V\omega^{-x_d})
   \pmatrix{1\cr1\cr\vdots\cr1\cr}
   ={1\over N}\sum_{\vec p}
   \phi(\vec p)e^{-2\pi i\vec p\cdot\vec x/L},
\end{eqnarray}
where $x_\mu=0$, 1, \dots, $L-1$, has a unique non-zero
element ($=1$) in the $i(x)$-th entry, where $i(x)$ is given by
eq.~(\ref{threexfour}). Since this collection of vectors forms a complete
basis, $\phi(\vec p)$ does too.

\listoftables           
\listoffigures          

\end{document}